\documentclass[aps,prl, twocolumn,superscriptaddress,letterpaper]{revtex4}

\usepackage{graphicx}
\usepackage{amssymb,amsmath}
\usepackage{float}
\usepackage{bm}

\usepackage{setspace}
\usepackage{parskip}
\usepackage{xcolor}
\definecolor{midnightblue}{cmyk}{1,1,0,0.1}
\definecolor{forestgreen}{cmyk}{0.76,0,0.26,0.5}

\usepackage{epsfig}

\usepackage{hyperref}
\hypersetup{
    bookmarks=true,         % show bookmarks bar?
    unicode=false,          % non-Latin characters in Acrobat bookmarks
    pdftoolbar=true,        % show Acrobat toolbar?
    pdfmenubar=true,        % show Acrobat menu?
    pdffitwindow=false,     % window fit to page when opened
    pdfstartview={FitH},    % fits the width of the page to the window
    pdftitle={My title},    % title
    pdfauthor={Author},     % author
    pdfsubject={Subject},   % subject of the document
    pdfcreator={Creator},   % creator of the document
    pdfproducer={Producer}, % producer of the document
    pdfkeywords={keyword1} {key2} {key3}, % list of keywords
    pdfnewwindow=true,      % links in new window
    colorlinks=true,       % false: boxed links; true: colored links
    linkcolor=midnightblue,          % color of internal links (change box color with linkbordercolor)
    citecolor=magenta,        % color of links to bibliography
    filecolor=midnightblue,      % color of file links
    urlcolor=midnightblue,          % color of external links
}

\begin{document}

\title{Valley Modulation and  Single-Edge Transport of Magnons in Staggered Kagome Ferromagnets}%
\author{Yuheng Xing}
\affiliation{NNU-SULI Thermal Energy Research Center (NSTER) \& Center for
Quantum Transport and Thermal Energy Science (CQTES), School of Physics and
Technology, Nanjing Normal University, Nanjing, China, 210023}

\author{Hao Chen}
\affiliation{NNU-SULI Thermal Energy Research Center (NSTER) \& Center for
Quantum Transport and Thermal Energy Science (CQTES), School of Physics and
Technology, Nanjing Normal University, Nanjing, China, 210023}

\author{Ning Xu}
\affiliation{Deparment of Physics, Yancheng Institute of Technology, Yancheng, China, 224051}

\author{Xiao Li}
\email{lixiao@njnu.edu.cn}
\affiliation{NNU-SULI Thermal Energy Research Center (NSTER) \& Center for
Quantum Transport and Thermal Energy Science (CQTES), School of Physics and
Technology, Nanjing Normal University, Nanjing, China, 210023}

\author{Lifa Zhang}
\email{phyzlf@njnu.edu.cn}
\affiliation{NNU-SULI Thermal Energy Research Center (NSTER) \& Center for
Quantum Transport and Thermal Energy Science (CQTES), School of Physics and
Technology, Nanjing Normal University, Nanjing, China, 210023}
%\date{25 May 2017}
\date{\today}% It is always \today,
% but any date may be explicitly specified

\begin{abstract}
Owing to its charge-free property, magnon is highly promising to achieve
dissipationless transport without Joule heating and thus potentially
applicable to energy-efficient devices. Moreover, a kagome lattice, as stacking layers of many magnon ferromagnets, also exhibits valley structure in quasiparticle spectra, which are likely to add a new dimension to magnon excitation. Here, we investigate valley magnon and associated valley modulation in a kagome lattice, with staggered exchange interaction and Dzyaloshinskii-Moriya interaction.  The staggered exchange interaction breaks spatial
inversion symmetry, leading to gapped degenerate valleys at $\pm K$ and consequent valley magnon Hall effect.  When the Dzyaloshinskii-Moriya interaction is further included, the valley degeneracy is lifted. As a result,  net magnon anomalous Hall effect and topological phase transition are realized. More interestingly, by tuning valley splitting and excitation frequency,  heat currents in the kagome strip can be localized at one edge  to achieve single-edge transport. Besides, for the kagmon lattice, the edge heat currents include local circulating contribution within triangular fine structure, together with currents flowing parallel to the edges. These findings give full play to spin and valley degrees of freedom and enrich energy-efficient magnonic device paradigms.
%transport can help to enrich our microscopic understanding of spin
%excitations for future magnonics applications.
\end{abstract}

% 85.75.-d Magnetoelectronics; spintronics: devices exploiting spin polarized
% 75.30.Ds Spin waves
% 75.47.-m Magnetotransport phenomena; materials for  magnetotransport
% 75.70.Ak Magnetic properties of monolayers and thin films
\maketitle

\textcolor{forestgreen}{\emph{\textsf{Introduction}.}}---
With Dzyaloshinskii-Moriya interaction (DMI)  that plays a role of vector potential like the Lorentz force \cite{dzya58, moriya60}, magnon Hall effect has been theoretically predicted \cite{katsura10, Hoogdalem13} and experimentally observed in magnetic insulators \cite{onose10}. The magnon Hall systems, also named topological magnon insulator \cite{zhang13}, are characterized by nonzero Chern numbers and topologically protected magnon edge states, similar to electronic topological insulators \cite{Hasan10,Qi11}. Quantum transport based on topological magnon edge states are highly promising to achieve dissipationless transport without ohmic loss \cite{Shindou13,Mook14,Mena14,chis15,Nakata17,LiYM18}. On the other hand, valley degree of freedom has been fully expressed in electronic band structure of transition-metal dichalcogenides \cite{Xiao07, Mak14, Lee16}, where valley carrier is selectively excited by chiral optical fields and measured by Berry-curvature-induced valley Hall current \cite{RHSacoto20}. Valley degeneracy is also tunable by applied magnetic field or magnetic proximity effect \cite{JSQ15}. Valleytronics, using valley degree of freedom as information carrier, has been a rising field  in condensed matter physics. The explorations of both topological magnons and valleytronics have not only conceptual importance in basic quantum physics,  but also application potential in advanced information technology \cite{Kruglyak10,Shindou13}.

\begin{figure}[h!]
\setlength{\belowcaptionskip}{-0.35 cm}
\centering
\includegraphics[width=8.0 cm]{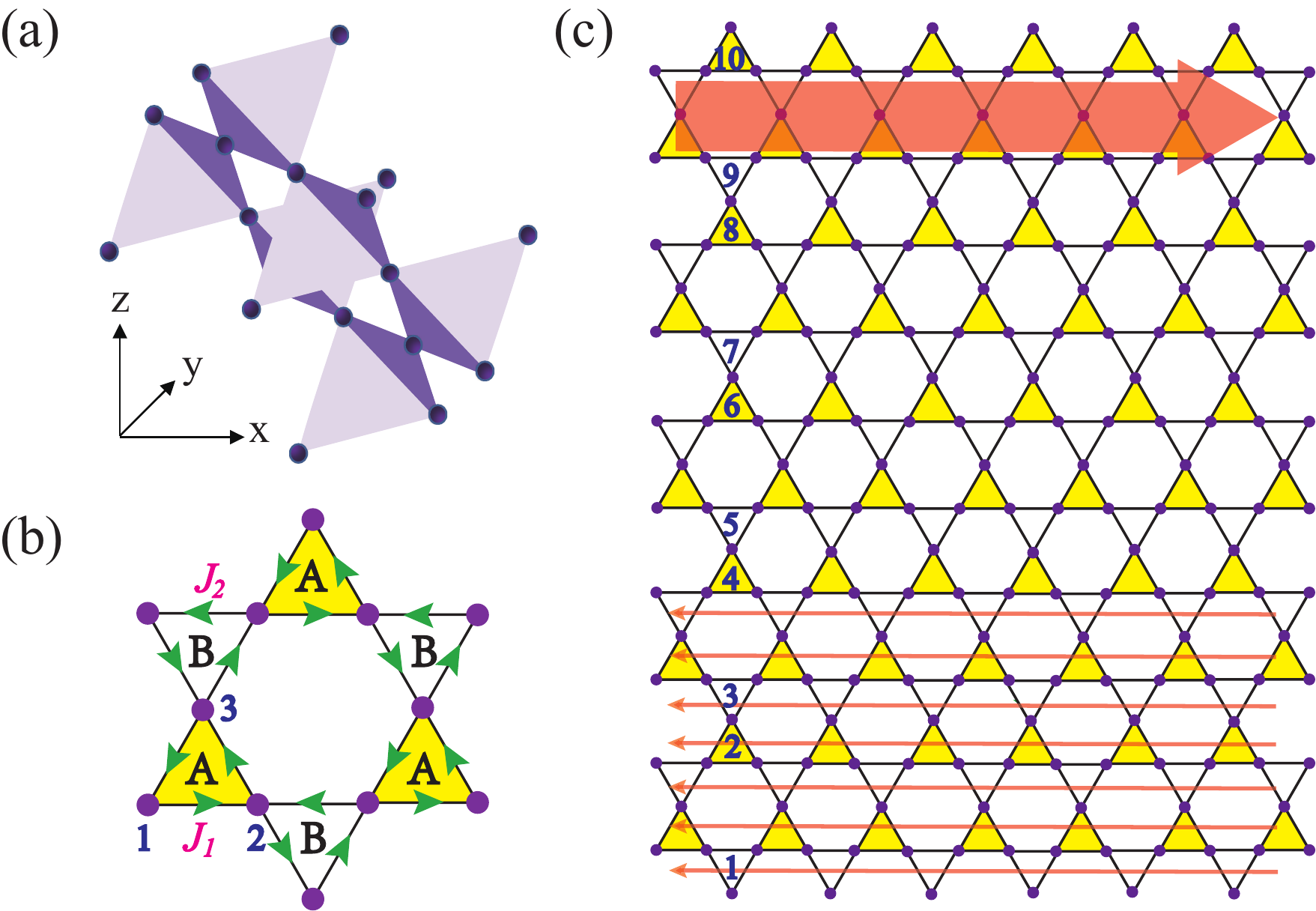}
\caption{Structures of the kagome lattice. (a) A pyrochlore  structure including kagome layers.
(b) A kagome lattice with staggered exchange interactions in neighboring A and B triangles. A unit cell includes three sites, denoted by 1-3. The arrows between nearest-neighboring sites denote the directions with DMI along $+z$. (c) A representive quasi-one-dimensional kagome strip, with ten triangles along its transversal direction.}
\label{fig1}
\end{figure}

Inequivalent energy valleys in magnon excitations, if exist, are expected to directly inherit merits from both magnons and valleys, and possible modulations of valley magnons are likely to further gives full play to spin and valley degrees of freedom. Compared with valley electrons, the studies of valley magnons, especially for valley modulation, are relatively few.  There are attractive questions, e.g. whether the valley degeneracy of magnons is tunable like its electronic counterpart  and correspondingly what transport behavior edge magnons exhibit.  Given that magnon Hall effect has been realized experimentally in pyrochlore structures with kagome layers \cite{Ideue12,onose10}  and valley physics is also present in the kagome lattice \cite{NLera19}, we take a two-dimensional ferromagnetic kagome lattice as a prototype to investigate valley modulation and band topology in magnon excitation.
In this Letter,  the effects from staggered exchange interaction (SEI) \cite{PHogl20} and DMI on valley magnons are taken into account.  The SEI in neighboring triangles of the kagome lattice (See Fig. 1), as well as DMI, creates  gaps at  $\pm K$ valleys and keep valley degeneracy. The combination of both interactions further lifts the valley degeneracy, and enables a transition from valley magnon Hall effect to net magnon anomalous Hall effect. The band exchange at valleys and associated topological phase transition also occur with varied interaction strengths.  Further considering topological edge transport,  besides the heat current flowing parallel to the edges,  local circulating current appears within triangles of the kagome lattice.  An asymmetric edge transport, mainly contributed by one edge, results from the valley splitting.  These intriguing features realize modulations of the valley degeneracy, band topology and associated edge transport in magnon excitation, which provides avenues for exploring energy-efficient device paradigms based on coupled spin and valley degrees of freedom.

\smallskip

\textcolor{forestgreen}{\emph{\textsf{Model}.}}---
Figure 1 (a) show the pyrochlore  structure formed by corner-sharing tetrahedra, which is the crystal structure of many magnon Hall ferromagnets, e.g. In$_2$Mn$_2$O$_7$ and Ho$_2$V$_2$O$_7$ \cite{Ideue12}.  Along crystallographic [111] direction, there is a stacking of parallel kagome layers and intermediate triangular layers. Given that the kagome lattice supports both  topological magnons and valley structure,  we focus on valley magnons in a kagome monolayer. A spin Hamiltonian of the lattice is given as \cite{dzya58, moriya60,Heisenberg28,bose94,onose10,zhang13,chis15,chern16, YSu17, A.R18}:
\begin{equation}\label{eq_ham}
\mathcal{H}\!=\!
-J_1\!\sum_{\langle {mn}\rangle \in \!A}\!\bm S_m\!\cdot \bm S_n
\!-\!J_2\!\sum_{\langle {mn}\rangle \in\!B}\!\bm S_m\!\cdot \bm S_n
\!+\!D\!\sum_{\langle {mn}\rangle}\!\bm \xi_{mn}\cdot\bm S_m\!\times \bm S_n .
%  - g\mu _B \bm B \cdot\sum_i {\bm S_i }
\end{equation}
Here,  $\bm S_{m,n}$  are spins on sites $m,n$, of which the nearest-neighboring interactions are considered.  The first two terms describe exchange interactions in neighboring  A  and B triangles of the kagome lattice, respectively, as illustrated in Fig. 1b, with  $J_{1,2}$ being corresponding strengths. The third term denotes DMI with a strength of $D$. $\bm \xi_{mn}=\pm \bm z$ when the vector pointing from the $n$-th site to $m$-th site is parallel and antiparallel to the arrow in Fig. 1b, respectively.  For $J_{1}=J_{2}$, the lattice has spatial inversion symmetry, while $J_{1} \neq J_{2}$ breaks the symmetry. Different $J_{1,2}$ can result from additional layers adjacent to the kagome monolayer in real materials \cite{PHogl20,Yin19}.

By the Holstein-Primakoff transformation \cite{holstein40} and the Fourier transformation, the magnon Hamiltonian in the momentum space reads,
\begin{eqnarray}
\label{Hk}
\begin{split}
\mathcal H(\bm{k})\!=\!
\left(
\begin{array}{ccc}
\!2(J_1\!+\!J_2)S                                        &\!  f_{12}(\bm k)                       & \!f_{31}^*(\bm k)      \\
\!f^*_{12}(\bm k)                               &\!2(J_1\!+\!J_2)S                            &  \!f_{23}(\bm k) \\
 \!f_{31}(\bm k)          &\!f^*_{23}(\bm k)                   &  \!2(J_1\!+\!J_2)S
\end{array}
\right)
\end{split}
\end{eqnarray}
where $f_{\alpha\beta}(\bm k)=\eta_1\text{exp}(-i\phi_{\alpha\beta})+\eta_2\text{exp}(i\phi_{\alpha\beta})$,
with $(\alpha, \beta)=(1,2),(2,3)$ and $(3,1)$. $\eta_{1,2}=-(J_{1,2}+iD)S$ and $\phi_{\alpha\beta}=\bm k \cdot (\bm r_\alpha-\bm r_\beta)$. $\bm r_\alpha$ is the coordinate of the $\alpha$-th site in a unit cell  with $\alpha=1,2,3$ [Fig. 1(b)]. Without loss of generality, the spin magnitude, $S$, and $J_1$ are set to $\frac{1}{2}$ and the unit of the energy, respectively. $J_2-J_1$ and $D$ are one order of magnitude smaller than $J_1$.
Band structures and magnon properties, e.g. magnon Berry curvature, are computed by solving eigenstates of the Hamiltonian \ref{Hk}.
More details of deriving the Hamiltonian and calculations can be found in Supporting Information (S.I hereafer).

\smallskip

\textcolor{forestgreen}{\emph{\textsf{Magnon bands and  topological properties}.}}---
Fig. 2 (a-d) shows magnon band structures of the kagome lattices, where the bands of the lattice with uniform exchange interaction ($J_1=J_2$) and vanishing DMI are taken as references. For the case,  the topmost band is completely flat, which touches with the middle band at the Brillouin zone center, $\Gamma$. The middle and lowest bands have Dirac-type dispersion in the vicinity of $\pm K$ points,  i.e.  vertices of the hexagonal Brillouin zone, and they touch at exactly $\pm K$, exhibiting degenerate, gapless valley structure.  The entire magnon excitation is thus gapless as well.

\begin{figure}[htbp]
\centering
\includegraphics[width=9 cm]{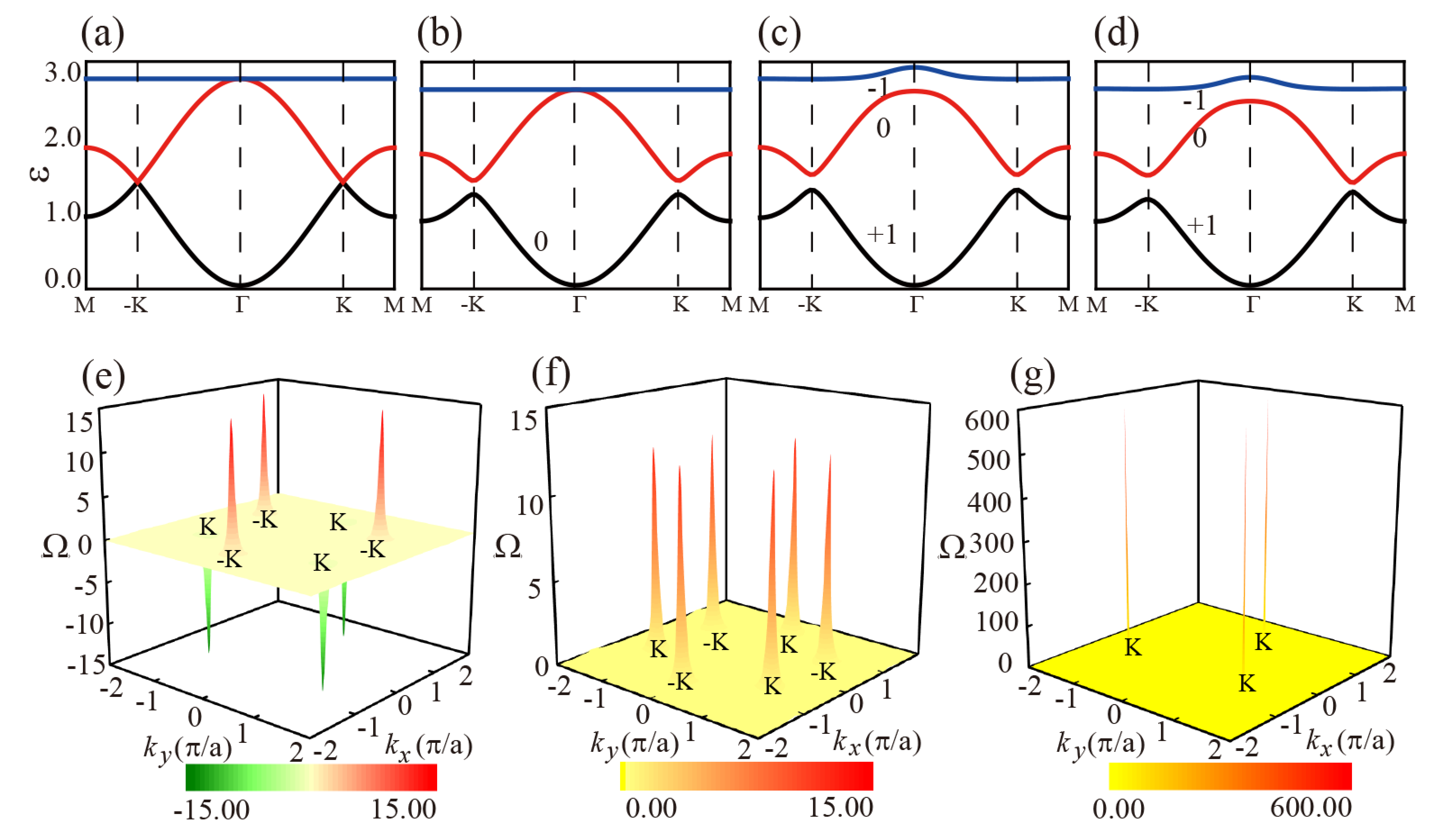}
\caption{Magnon dispersions and topological properties of kagome lattices. The band structures of the lattices (a) without SEI and DMI, (b) with SEI only, (c) with DMI only and  (d) with both interactions. The parameters $(J_2, D)=(1.0, 0.0), (0.9, 0.0), (1.0, 0.1)$ and $(0.9, 0.1)$ in (a-d), respectively, with $J_1=1.0 $ being the unit of energy. The Chern
number is labeled for each isolated band. (e-g) The $\Omega (\bm k)$ of the lowest band, using the parameters in (b-d), respectively. }
\label{fig2}
\end{figure}

With SEI considered,  band gaps open between the lower two bands at $\pm K$ valleys, while the flatness of the topmost band and the connection between the upper two bands at $\Gamma$ are well kept, as shown in Fig. 2(b). When nonzero DMI is introduced together with $J_1=J_2$, each band becomes isolated in Fig. 2(c), with band gap openings at both $\pm K$  and $\Gamma$. The topmost band is no longer flat, with a bulge at $\Gamma$. For the above two cases, the valley degeneracy is still present, with equal gaps at $\pm K$ valleys.

Taking into account both SEI and DMI,  besides isolated bands, the valley degeneracy is lifted, that is, magnon energies and gaps at $\pm K$ become unequal in Fig. 2(d). The valley splitting offers a energy window for realizing valley-selective magnon excitation by magnon waveguide.
The valley index corresponding to a large (or small) gap is determined by both the sign of DMI and relative magnitude of SEI. When the sign of $(J_1-J_2)\cdot D$ is positive, large and small gaps are localized at $K$ and $-K$ valleys, respectively. When the sign is negative, the relative magnitudes of the gaps are interchanged. 

The momentum-resolved Berry curvature, $\Omega (\bm k)$, and  associated Chern number of  magnon bands are further computed as indicators of topological quantum transport. They are shown in Figs. 2 (e-g) and labeled on each isolated band in Figs. 2 (b-d), respectively. When both SEI and DMI are absent,  $\Omega (\bm k)$ is zero for each $\bm k$ point without band crossing.  Therefore, there is no anomalous Hall transport induced by Berry curvature.

When SEI or DMI  is added, $\Omega (\bm k)$ becomes non-vanishing, and it has extrema at  $\pm K$ valleys for the lower two bands.  Although corresponding band structures in the neighborhood of  $\pm K$  valleys are similar [Fig. 2 (b-c)], the distributions of  $\Omega (\bm k)$ are distinct. For SEI, $\Omega (\bm k)$ has the same magnitude but opposite signs at two valleys. The Chern number, as integral quantity of $\Omega (\bm k)$, is thus zero for the isolated lowest band. Although the magnon band is topologically trivial,  opposite $\Omega (K)$ and  $\Omega (-K)$  endow valley magnons opposite anomalous velocities and consequently induce transversal Hall heat currents along opposite directions, under a longitudinal temperature gradient. It is a magnon version of valley Hall effect.
As for nonzero DMI, $\Omega (\bm k)$ is the same at $\pm K$ valleys, leading to a Chern number of 1 for the lowest band and associated topological edge transport. The edge transport will be discussed in the next section. 

For a staggered lattice with nonzero DMI, $\Omega (\bm k)$ at two valleys does not exhibit the same magnitude any longer, owing to the valley splitting.  As a result,  a net magnon anomalous Hall heat current is realized, no matter single valley or both valleys are excited by magnon waveguide, since anomalous Hall currents contributed by two valleys can't completely cancel with each other. This is in contrast to the lattice with only SEI where the net Hall current is absent. The net Hall heat current can be readily measured by induced transversal temperature difference, which is expected to be used as information carrier in advanced device paradigms. The distinct Berry curvatures in Fig. 2 (e-g) are associated with distinct massive terms of gapped Dirac states arising from SEI and DMI, which are analyzed in S.I. by an effective Dirac model at $\pm K$ valleys.

Moreover,  topological phase transitions can occur, depending on interaction strengths. The topological phase diagram is shown in S.I. as a function of $J_2$ and $D$.  For the isolated lowest band, the phase transition can be easily found from two limits in Fig. 2(b-c). The lowest band has the Chern number of 0 and 1, with only SEI and only DMI considered, respectively. For an existing SEI, the Chern number changes from 0 to 1 as  DMI is added and enhanced. In the process, the band gap at $K$ becomes small, closed and reopened, while the gap at $-K$ always increases. The band exchange at $K$ leads to  varied Chern number.  Therefore, the Chern number is determined by the relative magnitudes of SEI and DMI. Fig. 2(d) corresponds to a relatively large DMI, and its bands have the same Chern numbers with the ones in Fig. 2(c).
%Moreover, there are topological phase transitions, depending on interaction strengths. The phase diagrams of band topology for three band are shown in Fig. 2(h-j), respectively, as  functions of $D$  and $J_2$. In the phase diagrams, different regions correspond to different Chern numbers.  For the isolated lowest band, the phase transition can be easily found from two limits in Fig. 2(b-c). The lowest band has the Chern number of 0 and 1, with only SEI and only DMI considered, respectively. For an existing SEI, the Chern number will change from 0 to 1 as  DMI is added and increased. In the process, the band gap at $K$ become small, closed and reopened, while the gap at $-K$ is always increased. The band exchange  at $K$ leads to  varied Chern numbers.  Therefore, the Chern number is determined by the relative magnitude of  $J_2$ and $D$. Fig. 2(d) corresponds to a relatively large $D$ compared with $J_2$ and it has the same Chern number with that in Fig. 2(c) for each band.

\smallskip

\textcolor{forestgreen}{\emph{\textsf{Chiral magnon edge transport}.}}---
Considering the above significant modifications of valley and topological properties in the kagome monolayer, we further study magnon transport carried by topological edge states.   Given that topological edge states are associated with nonzero Chern number according to bulk-edge correspondence and nonvanishing DMI gives arise to nonzero Chern number, two cases, with only DMI considered and with both DMI and SEI, are focused on below. A nanostrip of the kagome lattice, similar to that in Fig. 1(c), is constructed with 40 triangles along its transversal direction and used in our calculation.

\begin{figure}[htbp]
\centering
\includegraphics[width=8.5cm]{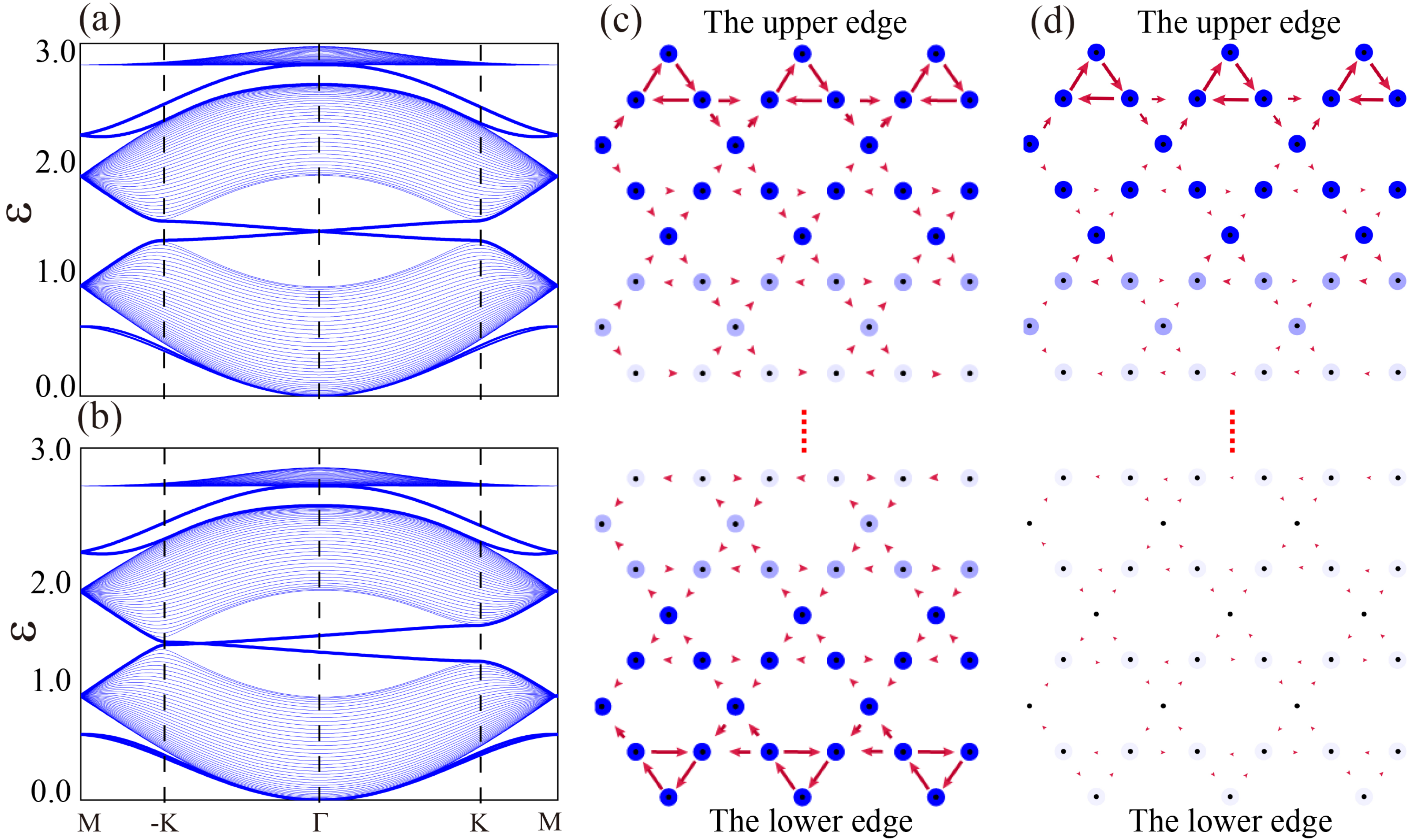}
\caption{Magnon properties of kagome strips.  The band structures of the strips (a) with DMI only and (b) with both interactions. %Their paremeters are the same with Figs. 2 (c) and (d), respectively.
(c) and (d) Corresponding local heat current and local density in the strips. The arrow and circle denote the current and density, respectively, with the size and color saturation being proportional to respective magnitude. %The arrow also gives the current direction. 
 }
\label{fig3}
\end{figure}

Figs. 3 (a) and (b) demonstrate magnon bands of the kagome strips with only DMI and with both interactions, respectively. Their choices of $J_{1,2}$ and $D$ are the same with Figs. 2 (c) and (d), respectively, which ensures that corresponding bulk bands of two strips have the same Chern numbers.  For both strips, there are indeed edge states within band gaps. The edge states within each gap are composed by one pair of gapless Dirac modes with opposite group velocities. % indicating topological characteristics and agreeing with calculated bulk Chern numbers that have a difference of 1 between neighboring bulk bands.
For edge modes between the upper two bulk bands,  the crossing point of the counterpropagating modes, i.e. the Dirac point, is located at the boundary of one-dimensional Brillouin zone, $M$. The differences between two strips are embodied in edge modes within the lower band gap, which connect $\pm K$ valleys. With only DMI, the edge dispersions along $\Gamma$ to $\pm K$  paths are symmetric with respect to $\Gamma$, and the Dirac point of edge modes is right at $\Gamma$. The degeneracy of edge modes is consistent with the valley degeneracy of bulk bands. With both interactions considered, the symmetry is broken and the Dirac point of edge modes moves away from $\Gamma$, due to the valley splitting.

We then construct a two-terminal device of the kagome strip to investigate topological edge heat transport within the lower gap, using non-equilibrium Green's function method \cite{Haug96}.
Two semi-infinite leads are added at left and right terminals of the strip, with temperatures of $T_L$ and $T_R$, respectively.  Equilibrium transport without the temperature gradient ($T_L=T_R$) is firstly considered, and non-equilibrium transport ($T_L\neq T_R$) will be also discussed in the end of the section.
The transmission coefficient of the edge transport is  calculated for the strips with only DMI and with both interactions. Since the transmission coefficient is determined by the number of edge modes,  it is always 1 within the band gap for the two strips with the same band topology.

In order to demonstrate spatial distribution of edge transport,  local magnon density, $\rho_n(\epsilon)$, at the site $n$ and local heat current, $j_{mn}(\epsilon)$, from the site $n$ to its nearest-neighboring site $m$ are computed for a given energy, $\epsilon$ \cite{zhang13, Li09}. Their distributions in the lattice are demonstrated in Figs. 3(c-d). For both strips, there are circulating currents flowing along three bonds of triangles in the kagome lattice.  The magnitudes of $j_{mn}$ along two oblique bonds are the same, but unequal to that along the horizontal bond. Therefore, a net forward or backward current along the horizontal bond, i.e. the edge direction of the strip, also exists, together with  circulating current. For each triangle, the circulating current, $j_C$, and the net horizontal current, $j_H$, are defined as local current along the oblique bond and current difference between the horizontal and oblique bonds, respectively, with the direction from left to right being positive direction. %Taking the A triangle in Fig. 1(b) for example, $j_C=j_{32}=j_{13}$ and  $j_H=j_{21}-j_{13}$.
 $j_H$ and  $j_C$ are shown in Fig. 4 for each triangle along the transversal direction of the strip, while the longitudinal direction keeps translational symmetry.  For the triangles with sizable local currents, both  $j_{H}$ and $j_{C}$ are non-vanishing, and $j_{C}$ is several times larger than $j_{H}$.   The nonzero $j_H$ is determined by the band topology and it takes responsibility for topological chiral edge transport, while nonzero $j_C$ is magnon characteristic of the kagome lattice with triangular fine structure. %The two kinds of local currents are present for both cases with DMI, no matter the magnitude of SEI.
%The circulating current results from the chiral DM interaction and local triangular structure [J$_{1,2}$ do not?].

\begin{figure}[htbp]
\setlength{\belowcaptionskip}{-0.35 cm}
\centering
\includegraphics[width=8 cm]{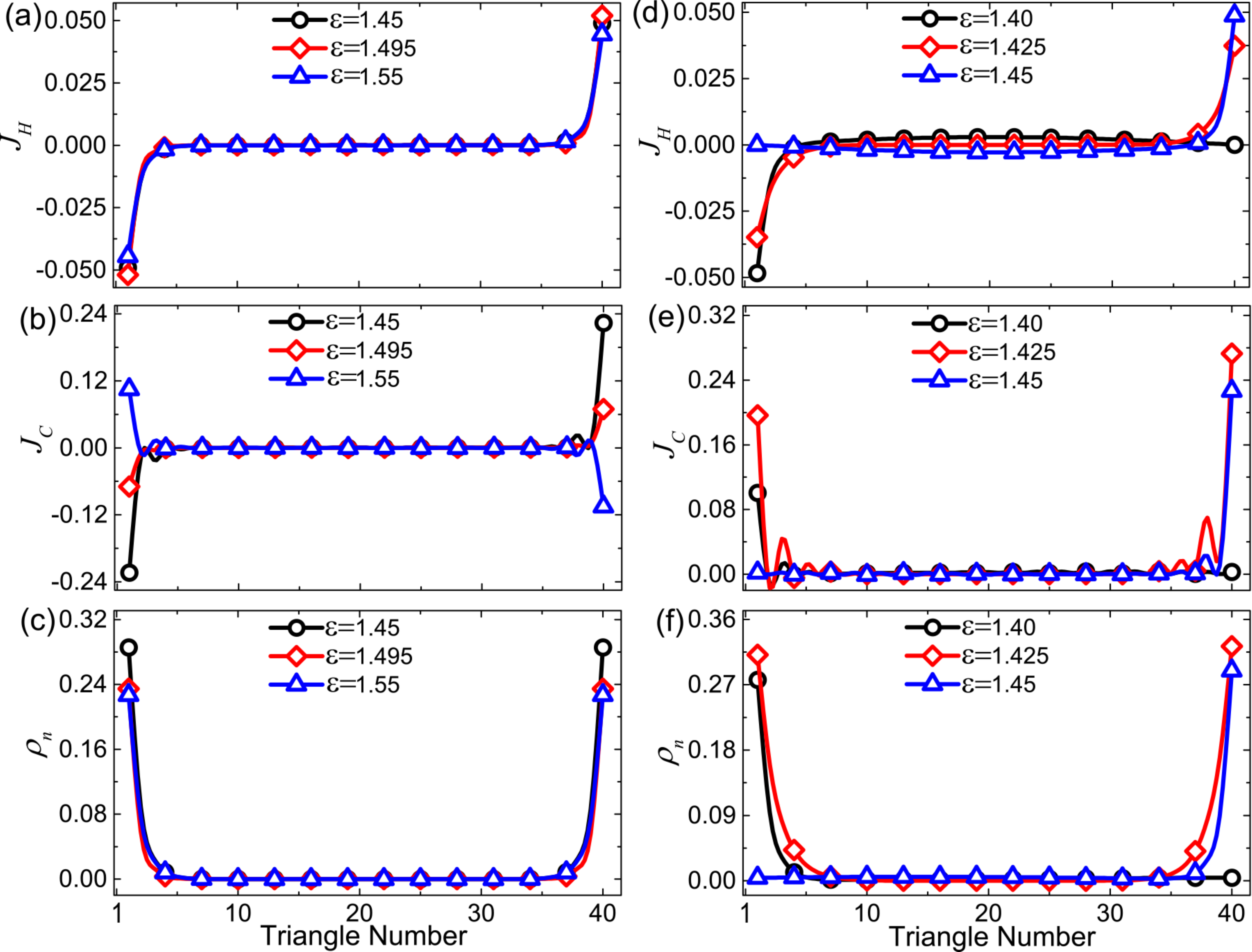}
\caption {The evolutions of local horizontal current, circulating current and density along the transversal direction of the strip.  (a-c) These quantities in a strip with only DMI. (d-f)  correspond to a strip with both interactions.}
\label{fig4}
\end{figure}

Looking at evolutions of local densities and heat currents along the transversal direction of the strip, there are distinct distributions of edge modes in Figs. 3(c-d).  With only DMI included, edge modes are symmetrically localized at two edges of the strip. That is, $j_H$, $j_C$ and $\rho_n$ all decay quickly when moving away from the edges, with the same magnitudes and decay rates at two edges [Figs. 4(a-c)]. For both $j_H$ and $j_C$, there is also a sign change between opposite edges.  Opposite $j_H$  correspond to counterpropagating edge modes in Fig. 3(a), giving rise to the same chirality. %As for  $j_C$,  it flows clockwisely and anticlockwisely [?] within the A triangles at two edges, respectively, while the direction becomes opposite for the B triangles. 
 Moreover,  comparing various energies within the gap, the symmetric distribution of the edge transport is well kept, while the sign of $j_C$ is tunable with the varied energy and there are also quantitative changes of $j_H$ and $\rho_n$.

For the strip with both SEI and DMI,  local heat currents and densities appear with obvious asymmetric distributions along the transversal direction when the magnon energy is close to bulk band edge. For an energy, $\epsilon=1.45$, near the upper band,  sizable currents and densities are well localized at the upper edge of the strip, while they disappear in the lower half of the strip, as demonstrated by the distributions in Fig. 3(d) and blue curves in Figs. 4(d-f). Instead, a weak background $j_{H}$ spreads transversely across the entire strip, and it is opposite to $j_{H}$ at the upper edge [See Figs. 4(d)], ensuring that the total chiral current is zero at equilibrium. The asymmetric distribution exhibits characteristics of single-edge transport. It is because the valley splitting makes two counterpropagating modes at the chosen energy have distinct energy differences with respect to bulk band and consequent distinct spatial localizations. Moreover, when the chosen energy is near the lower bulk band ($\epsilon=1.40$), the single-edge transport is shifted to the lower edge, with the horizontal  currents flowing in the opposite direction.  When the energy is close to the midgap ($\epsilon=1.425$), edge modes are far away from bulk bands in energy and consequently localized at both edges with a weak asymmetry. The evolutions of $J_{H,C}$ and $\rho_n$ at outermost triangles of two edges with the varied energy are also given in S.I. to further demonstrate frequency-tunable asymmetric transport. %Three distinct energy regions are identifed. At Regions I and III away from the Dirac point, the contribution from the upper edge and lower edge to the transport is negligible, respectively, giving rise to single-edge transport. At Region II including the Dirac point, the contributions from two edges  continually varies with the energy and asymmetric distribution also exists except for the Dirac point.

Moreover, the chiral edge transport is influenced by the signs of $(J_1-J_2)$ and $D$, similar to valley splitting of bulk band.  Opposite $J_1-J_2$ leads to  single-edge transports localized at opposite edges, with opposite current directions. The chirality of edge transport is thus invariant. In contrast, opposite $D$ gives rise to a direction reversal of the current at the same edge, leading to edge transports with opposite chiralities. The change of the chirality results from the reversal of bulk Chern numbers by reversing $D$, while opposite $J_1-J_2$ gives the same band topology and chirality. Moreover, the modification of edge transport is expected to be detected experimentally, e.g. by spatial-resolved magnon temperature measurement \cite{MAgrawal13} %using Brillouin light scattering spectroscopy  %[PRL 111, 107204 (2013)]
or magneto-optical Kerr effect \cite{DAAllwood03}. The tunable topological edge transport has potential use in information encoding and manipulation at the microscopic level.
%Moreover, similar to opposite valley splittings determined by the sign of $(J_1-J_2)\cdot D$,  opposite $J_1-J_2$ leads to a single-edge transport with opposite current direction and at the opposite edge, while opposite $D$ only gives rise to a reversal of the current direction. The edge distribution modification from the valley splitting is expected to be detected experimentally, e.g. by spatial-resolved magnon temperature measurement \cite{MAgrawal13} %using Brillouin light scattering spectroscopy  %[PRL 111, 107204 (2013)]
%or magneto-optical Kerr effect \cite{DAAllwood03}. The tunable topological edge transport has potential use in information encoding and manipulation at the microscopic level.

For completeness, we also computed topological edge transport when  two leads have different temperatures. % as shown in Fig. SX of S.I.
Even if only DMI is present, there are asymmetric currents and densities by comparing two edges. The edge current, flowing from heat lead to cold lead, is enhanced, while the opposite current is weakened, in contrast to the symmetry distribution at equilibrium. The above changes also apply to the strip with both SEI and DMI.

\textcolor{forestgreen}{\emph{\textsf{Conclusion}.}}---In summary, we studied valley magnons in a kagome lattice. Valley degeneracy and band topology are tunable by introducing SEI and DMI. As a result, valley magnon Hall effect and topological edge transport are expected to be realized.  Topological edge transport includes local circulating currents due to fine triangular structure of the kagome lattice and currents flowing along the edge direction. These currents can be localized at one edge to achieve single-edge transport by tuning valley splitting and frequency. The valley modulation of magnons and associated asymmetric transport add  different dimensions to the exploration of unique device paradigms based on  coupled spin and valley degrees of freedom.

\bigskip

%\vspace{0.2in}
\textcolor{forestgreen}{\emph{\textsf{Acknowledgments}.}}---We thank Qian Niu and Jian-Sheng Wang for helpful discussions. The work was supported by the National Natural Science Foundation of China (Nos. 11890703, 11975125, 11904173) and  MOST (Nos. 2017YFA0303500).


\begin{thebibliography}{99}
\bibitem{dzya58}  I. Dzyaloshinskii, J. Phys. Chem. Solids \textbf{4}, 241 (1958).
\bibitem{moriya60} T. Moriya, Phys. Rev. \textbf{120}, 91 (1960).
\bibitem{Hoogdalem13} K. A. van Hoogdalem, Y. Tserkovnyak, D. Loss, Phys. Rev. B \textbf{87}, 024402 (2013).
\bibitem{katsura10} H. Katsura, N. Nagaosa, P. A. Lee,  Phys. Rev. Lett. \textbf{104}, 066403 (2010).
\bibitem{onose10} Y. Onose, T. Ideue, H. Katsura, Y. Shiomi, N. Nagaosa, Y. Tokura, Science  \textbf{329}, 297 (2010).
\bibitem{zhang13} L. Zhang, J. Ren, J. S. Wang, B. Li, Phys. Rev. B \textbf{87}, 144101 (2013).
\bibitem{Hasan10} M. Hasan, C. Kane, Rev. Mod. Phys. \textbf{82}, 3045 (2010).
\bibitem{Qi11} X. Qi, S. Zhang, Rev. Mod. Phys. \textbf{83}, 1057 (2011).
\bibitem{Mena14} M. Mena, R. Perry, T. Perring, M. Le, S. Guerrero, M. Storni, D. Adroja, Ch. Ruegg, D. McMorrow, Phys. Rev. Lett. \textbf{113}, 047202 (2014).
\bibitem{chis15} R. Chisnell, J. S. Helton, D. E. Freedman, D. K. Singh, R. I. Bewley, D. G. Nocera, Y. S. Lee, Phys. Rev. Lett. \textbf{115}, 147201 (2015).
\bibitem{Owerre17} S. A. Owerre, J. Phys. Commun. \textbf{1}, 025007 (2017).
\bibitem{Shindou13} R. Shindou, R. Matsumoto, S. Murakami, J. I. Ohe, Phys. Rev. B \textbf{87}, 174427 (2013).
\bibitem{Mook14} A. Mook, J. Henk, I. Mertig, Phys. Rev. B \textbf{90}, 024412 (2014).
\bibitem{Nakata17} K. Nakata, S. K. Kim, J. Klinovaja, D. Loss, Phys. Rev. B \textbf{96}, 224414 (2017).
\bibitem{LiYM18} Y. M. Li, J. Xiao, K. Chang, Nano Lett. \textbf{18}, 3032-3037 (2018).
\bibitem{Xiao07} D. Xiao, W. Yao, Q. Niu, Phys. Rev. Lett. \textbf{99}, 236809 (2007).
\bibitem{Mak14} K. F. Mak, K. L. McGill, J. Park, P. L. McEuen, Science \textbf{344}, 1489 (2014).
\bibitem{Lee16} J. Lee, K. F. Mak, J. Shan, Nat. Nano. \textbf{11}, 421 (2016).
\bibitem{RHSacoto20} R. H. Sacoto, R. I. Gonzalez, E. E. Vogel, et al, arXiv preprint arXiv:2001.11934v1, (2020).
\bibitem{JSQ15} J. S. Qi, X. Li, Q. Niu, J. Feng, Phys. Rev. B \textbf{92}, 121403 (2015).
\bibitem{Kruglyak10} V. V. Kruglyak, S. O. Demokritov, D. Grundler, J. Phys. D \textbf{43}, 264001 (2010).
\bibitem{Ideue12} T. Ideue, Y. Onose, H. Katsura, Y. Shiomi, S. Ishiwata, N. Nagaosa, Y. Tokura, Phys. Rev. B \textbf{85}, 134411 (2012).
\bibitem{NLera19} N. Lera, D. Torrent, P. S. Jose, J. Christensen, J. V. Alvarez, Phys. Rev. B \textbf{99}, 134102 (2019).
\bibitem{PHogl20} P. Hogl, T. Frank, K. Zollner, D. Kochan, M. Gmitra, and J. Fabian, Phys. Rev. Lett. \textbf{124}, 136403 (2020).
\bibitem{YSu17}Y. Su, X. S. Wang, X. R. Wang, Phys. Rev. B  \textbf{95}, 224403 (2017).
\bibitem{Heisenberg28} W. Heisenberg, Z. Phys. \textbf{49}, 619 (1928).
\bibitem{bose94} I. Bose and U. Bhaumik, J. Phys.: Condens.Matter \textbf{6}, 10617 (1994).
\bibitem{chern16} A. L. Chernyshev and P. A. Maksimov, Phys. Rev. Lett. \textbf{117}, 187203 (2016).
\bibitem{A.R18} A. Rckriegel, A. Brataas, R. A. Duine, Phys. Rev. B \textbf{97}, 081106 (2018).
\bibitem{Yin19} J. Yin, S. Zhang, G. Zhang, et al, Nature Physics \textbf{443}, 15 (2019).
\bibitem{holstein40} T. Holstein, H. Primakoff, Phys. Rev. \textbf {58}, 1098 (1940).
\bibitem{Haug96} H. Haug and A. P. Jauho, Quantum Kinetics in Transport and Optics of Semiconductors (Springer, Berlin, 1996).
\bibitem{Li09} J. Li, T. C. A. Yeung, C. H. Kam, X. Zhao, Q. H. Chen, Y. Peng, and C. Q. Sun, J. Appl. Phys. \textbf{106}, 054312 (2009).
\bibitem{MAgrawal13} M. Agrawal, V. I. Vasyuchka, A. A. Serga, A. D. Karenowska, G. A. Melkov, B. Hillebrands, Phys. Rev. Lett. \textbf{111}, 107204 (2013).
\bibitem{DAAllwood03} D. A. Allwood, G. Xiong, M. D. Cooke, et al. J. Phys. D: Appl. Phys., \textbf {36}, 2175 (2003).






\end{thebibliography}
\end{document}